\documentclass[reprint, amsmath, amssymb, aps, showpacs, pra]{revtex4-2}

\usepackage{graphicx}
\usepackage{dcolumn}
\usepackage{bm}
\usepackage{hyperref}
\usepackage{physics}

\begin{document}
\title{Mitigating Heating of Degenerate Fermions in a Ring-Dimple Atomic Trap}

\author{Daniel G. Allman}
\email{Daniel.G.Allman.GR@Dartmouth.edu}
\author{Parth Sabharwal}
\author{Kevin C. Wright}

\affiliation{Department of Physics and Astronomy, Dartmouth College, 6127 Wilder Laboratory, Hanover NH 03766, USA}

\begin{abstract}
We report on the impact of the extended geometry of a ring-dimple trap on particle loss heating of a degenerate Fermi gas. When the Fermi level is slightly greater than the depth of the dimple and a non-degenerate ``halo'' is present, the overall heating rate is reduced relative to the case of a bare ring. We find that the experimentally measured heating rates for the overfilled dimple are in good agreement with a model of the hole-induced heating caused by background gas collisions. This suppression of the heating rate can be helpful for experimental studies of fermionic superfluids in the weak pairing limit, where achieving and maintaining low temperatures over long time scales is essential.
\end{abstract}

\maketitle

\textit{Introduction.} Superfluidity in weakly-interacting Bose-Einstein condensates is relatively well-understood, but fermionic superfluidity is much more complex, especially in superfluid phases that involve unconventional pairing mechanisms. Experimental studies of fermionic superfluids with weak BCS or exotic pairing require low temperatures to avoid pair-breaking by thermal excitations, but achieving sufficiently low temperatures in ultracold atomic systems is difficult because of limits on cooling efficiency, and the fact that deeply degenerate fermionic systems are extremely sensitive to heating caused by losses. Even if a system is prepared at an arbitrarily low temperature, collisions with background gas molecules inevitably knock particles out of the Fermi sea and heat the system above the critical temperature, on a time scale that depends on the collision rate and the thermodynamic properties of the system. This potential problem was recognized during early work on ultracold Fermi gases~\cite{Timmermans2001}, and the practical limits on experimentally achievable temperatures due to this mechanism have been previously considered for uniform and harmonically trapped Fermi gases, and for harmonically trapped Fermi-Bose mixtures~\cite{CoteOnofrioTimmermans2005}.

Hole-heating rates can be especially important for experiments aimed at probing low-energy long-wavelength dynamics in matter-wave circuits. The longest timescales for such experiments are set by the period of the lowest quantized circulation state, which can easily be seconds for larger systems. This paper considers the effects of hole-heating of fermions in a ring-shaped optical trap, and highlights the impact of the extended trap geometry on the thermodynamics of the system. We specifically consider the effects of hole heating on experiments performed on an equal spin mixture of $^6$Li in the BCS limit where pairing energies decreases exponentially with interaction strength. One important result of this analysis is that heating rates in fermionic matter-wave circuits can by reduced by using a dilute background population of atoms as a heat sink. We note too that although dimple traps have been routinely employed to increase the degeneracy of Fermi or Bose ensembles~\cite{Zurn2012,Duarte2015,Guttridge2019,StamperKurn1998,Stellmer2013,Viverit2001,Schuck2011}, little known work in an ultracold atoms setting has demonstrated the utility of a dimple trap to preserve large degeneracies in the presence of loss.

\textit{Fermionic Atoms in a Ring-Dimple Optical Trap.} Past experiments with atomic Bose-Einstein condensates (BEC) in ring-shaped or other multiply-connected trap potentials have used a wide range of magnetic and optical trapping techniques~\cite{GuptaBose-EinsteinPRL05, ArnoldPRA2006, RyuObservationPRL07, HendersonExperimentalNJP09, Bruce2011, SherlockTimePRA11, RamanathanSuperflowPRL11,  BeattiePersistentPRL13, Neely2013,  Navez2016, DeGoerDeHerve2021}. Experiments with ultracold Fermi gases generally require control of interactions using magnetic Feshbach resonances, making it necessary to use all-optical trapping techniques. Ring-shaped optical traps can be created using various combinations of red and/or blue-detuned beams, and recent experiments with rings of ultracold fermions have explored several of these possibilities already~\cite{CaiPersistentPRL2022, DelPaceImprinting}. The conclusions of this paper can be applied rather generally to these types of optical potentials, but we will focus our discussion on the red-detuned ring traps we used in our first experiments with rings of ultracold fermions, which we found to have some interesting and helpful features. 
\begin{figure}[t]
	\centering
 	\includegraphics[width=\columnwidth]{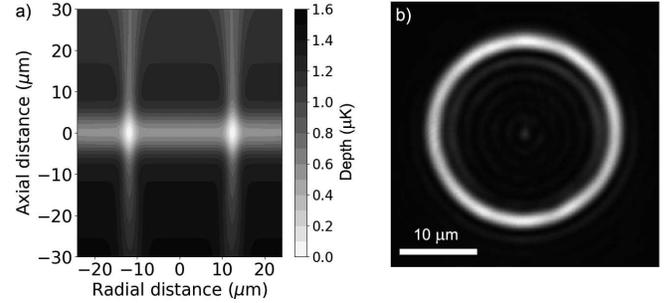}
	\caption{a) Side-view cross-section of a red-detuned optical trap potential used in our experiments with rings of fermionic atoms. This model includes the effects of the ring beam, the sheet beam and gravity. b) Intensity profile of the vertically propagating red-detuned ring-pattern beam modeled in (a).}
	\label{fig:ring beams profile}
\end{figure}

Optical ring traps typically employ at least two independent laser fields, one providing mainly vertical confinement, and the other radial. In our experiments with $^6$Li so far, the main vertical confinement was provided by a red-detuned (1064 nm) horizontally-propagating asymmetric Gaussian beam. In most of our experiments, the radial confinement was provided by a red-detuned (780 nm) vertically propagating laser shaped into a ring-pattern beam (Fig.~\ref{fig:ring beams profile}). This overall red-detuned beam configuration is similar to those used in many previous experiments with ring-shaped Bose-Einstein condensates~\cite{RamanathanSuperflowPRL11}. If the chemical potential of a quantum gas is sufficiently small compared to the depth of the ring-dimple, the atoms will be localized to the ring potential minimum and it is reasonable to treat the transverse confinement as approximately harmonic about the minimum. It is more straightforward to analytically calculate the chemical potential and other important properties of the system when this approximation is valid.

\begin{figure}[t]
	\centering
	\includegraphics[width=0.955\columnwidth]{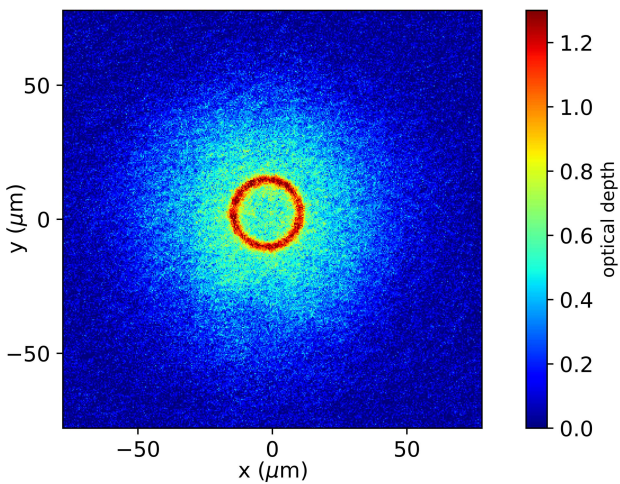}
	\caption{Density distribution of $^6$Li atoms in our trap when the Fermi level is around 0.1 $\mu$K larger than the depth of the dimple created by the ring-pattern beam. The figure shows the average of 10 in-situ absorption images taken at a magnetic field of $100$ mT. Both the ring-shaped region of increased density and the dilute halo are clearly visible.}
	\label{fig:atoms_in_trap}
\end{figure}

We had expected to conduct our first experiments with rings of ultracold fermions in this harmonic limit, but were surprised to see evidence that heating rates were lower when the Fermi level was high enough that the atoms filled the ring-shaped region of lowest potential and spilled over into the shallow extended potential created by the sheet beam, as shown in Fig.~\ref{fig:atoms_in_trap}. This dilute ``halo'' of atoms typically contained more than two-thirds of the total atom population, and played a crucial role in the thermodynamics of the system in our experiments. The harmonic approximation is clearly not valid for this situation, so we used a numerical 3D model of the potential to estimate the relevant thermodynamic properties of our system. 

\begin{figure}[t]
	\centering
	\includegraphics[width=\columnwidth]{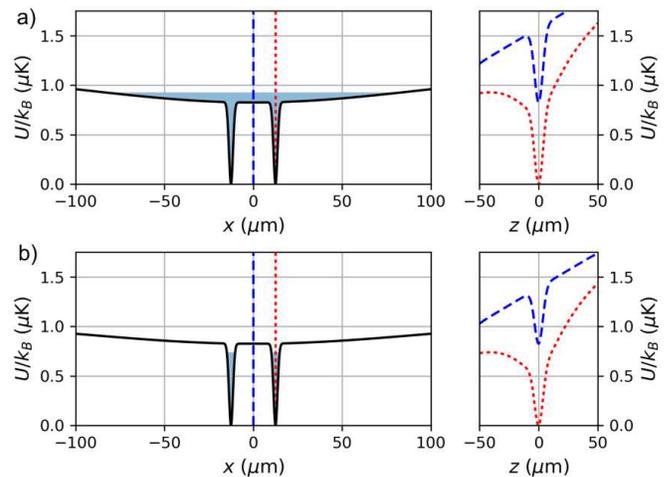}
	\caption{Potential energy slices of our combined trap (gravity included). a) 40 mW and b) 30 mW sheet power. Left plots show the radial trap profiles at $z=0$ while the right plots show the vertical profiles for two different radii. The blue (dashed) line is the vertical cut along $r=0$ and the red (dotted) line is along $r=r_0$. The shaded regions in the radial profiles (a,b) indicate where the potential energy is below the evaporation depth $V_{evap}$.}
	\label{fig:shallow_and_moderate_potentials}
\end{figure}

To compute the Fermi energy for atoms in this extended ring-dimple potential, we used a semi-classical model to obtain the total (spin up \textit{and} down) density of states, $g_{3D}(E)$, for a fully 3D model of the trap ($V(\textbf{r})$) that included the sheet beam, ring beam, and gravity:
\begin{equation}\label{3DDos}
	g_{3D}(E)=\frac{8\pi m}{(2\pi\hbar)^3}\int_{V(\textbf{r})\leq E}d^3r\sqrt{2m[E-V(\textbf{r})]}
\end{equation}
We then used the defining relation $N=\int_0^{E_F}g_{3D}(E)dE$ to numerically compute the Fermi energy $E_F(N)$, setting $E=0$ at the ring potential minimum. We modeled the ring beam as having an average radius of 12.5 $\mu$m and a transverse Gaussian profile with a radial $1/e^2$ half-width of 2.2(1) $\mu$m, in the plane of the sheet beam. Vertical trapping forces from this tightly focused ring were non-negligible, and so we found the through-focus intensity profile by numerically propagating the beam using the angular spectrum method ~\cite{goodman2005introduction} to obtain its full 3D profile. We modeled the sheet beam as having an asymmetric Gaussian profile with a horizontal waist of 290 $\mu$m and a vertical waist of 7 $\mu$m.

It turned out to be crucial to include the effects of gravity in the numerical calculation of the density of states. While gravity's effect on the exact value of the density of states at a given energy is small, the gradient due to gravity weakens the vertical confinement of atoms more substantially near the ring dimple region. This is conveniently visualized by plotting vertical cuts of the potential energy at radii near the ring radius $r_0=12.5\ \mu$m. These cuts each have a local maximum at some $z<0$ and linearly fall away to $-\infty$ for $z\ll 0$ due to gravity. The smallest of these maxima lies on the cut along $r=r_0$, as shown in Fig.~ \ref{fig:shallow_and_moderate_potentials}. Its potential energy sets the ``evaporation depth" $V_{evap}$ of the trap. Atoms with energy greater than this evaporation depth may overcome this ``lip" and fall out the bottom of the trap, and thus states with $E>V_{evap}$ should carry zero weight insofar as equilibrium thermodynamic quantities are concerned. We therefore multiply equation \eqref{3DDos} with the step function $\Theta(V_{evap}-E)$, which in turn has the dramatic effect of placing upper bounds on the allowed atom number and internal energy. 

As we will discuss below, there are additional subtleties in addressing the states with $E>V_{evap}$ that may remain bound to the trap via conservation laws that prevent escape through the evaporation channels near the ring dimple. The equilibrium configuration should not include these ``quasi-bound" orbits, but the relaxation dynamics may depend on them.

\textit{Fermi hole heating: Theory.}
In the BCS limit, the atomic trap lifetime is limited by the inelastic scattering rate with background particles. In a single background scattering event, a particle in the trap is ejected from the Fermi sea, leaving a hole behind. Assuming the subsequent relaxation dynamics does not eject any additional particles, the temperature increases slightly. For uniform one-body loss with lifetime $\tau_L$, the single-particle populations in state $|k\rangle$ and eigen-energy $\epsilon_k$ evolve according to $\dot{n}_k=-n_k/{\tau_L}$ ($k$ is a set of good single-particle quantum numbers for the inhomogeneous trap). The total atom number $N=\sum_k n_k$ and internal energy $U=\sum_k n_k \epsilon_k$ subsequently evolve as $\dot{N}=-N/\tau_L$ and $\dot{U}=-U/\tau_L$, respectively, where the single-particle loss equation was used. We note that the populations $n_k$ need not be thermally distributed. There are several equivalent methods of deriving the heating rate associated with this loss. Perhaps the most insightful method relies on the observation that the internal energy per particle $u\equiv U/N$ is a conserved quantity. Interestingly, this is true even during the elastic collisions that return the system from a non-equilibrium state to equilibrium after a hole is created. This fact implies that one can, at all times, meaningfully associate an effective temperature to the ensemble as if it were in equilibrium at the same energy and atom number. In our system, the thermodynamic variables used to describe the internal energy per particle $u$ are atom number $N$, temperature $T$, and a set of trap parameters which we call $\mathcal{V}$. We note that the \textit{only} thermodynamic role that $\mathcal{V}$ plays is in setting the energy scales for the single particle energy spectrum, which is fixed for the measurements performed in this paper as we are not varying the trap. The reversible mechanical work associated to trap deformations is therefore set to zero. We thus treat $N$, $U$ and $T$ as the only time-varying quantities under one-body loss, with the evolution $u(t)=u_0$ and $N(t)=N_0e^{-t/\tau_L}$ known, and that of $T(t)$ unknown. We can study the evolution $T(t)$ in a grand canonical picture, where a time-dependent chemical potential $\mu(t)$ is introduced and whose role is to fix $N(t)$ at each instant in time. We thus solve two equations
\begin{equation}\label{u0}
\frac{U(t)}{N(t)}=u_0=\frac{\int dE\ g(E)Ef[E;\mu(t),T(t)]}{\int dE\ g(E)f[E;\mu(t),T(t)]}
\end{equation}
and 
\begin{equation}\label{N}
N(t)=N_0e^{-t/\tau_L}=\int dE\ g(E)f[E;\mu(t),T(t)]
\end{equation}
for the two unknowns $T(t)$ and $\mu(t)$, where $f(E;\mu,T)=\{\exp[(E-\mu)/k_BT]+1\}^{-1}$ is the usual Fermi-Dirac distribution function and $g(E)$ the 3D density of states. By taking a time derivative of \eqref{u0} and utilizing \eqref{N}, it is possible to show that the evolution is equivalent to a differential equation governing the temperature dynamics. This is easier to demonstrate, however, by simply differentiating the internal energy function $U(N,T)$ with respect to time:
\begin{equation}\label{Tdot}
\dot{U}=\dot{N}(\partial U/\partial N)_T + \dot{T}(\partial U/\partial T)_N
\end{equation}
We then use the first law of thermodynamics $dU=TdS+\mu dN=T[dN(\partial S/\partial N)_T+dT(\partial S/\partial T)_N]+\mu dN$ to compute $(\partial U/\partial N)_T=\mu + T(\partial S/\partial N)_T$. Next, the Maxwell relation $(\partial S/\partial N)_T=-(\partial \mu/\partial T)_N$ is used to write $(\partial U/\partial N)_T=\mu -T(\partial \mu/\partial T)_N$. Finally, identifying the heat capacity at constant atom number $C_N=(\partial U/\partial T)_N$, we solve for the temperature derivative in \eqref{Tdot}:
\begin{equation}\label{heatingraterevised}
\dot{T}=-\frac{T^2\left(\frac{\partial}{\partial T}\frac{\mu}{T}\right)_N+u_0}{\tau_{L}c_N}
\end{equation}
with $c_N\equiv C_N/N$ and the time-dependent forms for $N(t)$ and $U(t)$ were used. This expression is in fact an extension of equation 5 in \cite{Timmermans2001}, which was derived using energy balance considerations, to arbitrary temperatures and inhomogeneous traps.

We emphasize now the role the halo plays in maintaining low temperatures for long periods of time. First, the large density of states offered by the broad sheet helps fermions disperse external energy imparted into the system into the closely spaced energy levels. Essentially, the dilute halo has a larger specific heat than the deeply-degenerate ring and can serve as an efficient heat sink, lowering the overall heating rate. Secondly, the halo acts as a particle reservoir for the ring-dimple, since the global chemical potential is only weakly dependent on the atom number when a substantial halo is present. Intuitively, any atom ejected from the ring-dimple can be ``replenished" by an atom in the halo. This in turn retains large densities in the ring-dimple region for longer periods of time. Combined, these two effects help maintain a deeply degenerate Fermi gas, especially in the ring-dimple region, for times exceeding the trap lifetime. 

In comparison, experiments performed in a ``bare" ring, i.e without a halo present, are likely to suffer from unacceptable heating rates. This may become particularly apparent in experiments utilizing a blue-detuned, repulsive ring beam, where the halo would typically be absent or separated from the superfluid, unless the potential is carefully tailored to make this possible. Blue-detuned traps have advantageous characteristics for some experiments, but the limits imposed by hole-heating will be a greater problem for experiments requiring many seconds to perform.

\textit{Fermi Hole Heating: Experiment.} 
The vacuum-limited lifetime of atoms in our glass cell experimental chamber is at least one minute. To help ensure we could clearly distinguish the effects of hole heating from the effects of slow technical drifts in experimental conditions, we deliberately reduced the lifetime to 25 seconds by shutting off the ion pumps attached to the 3DMOT vacuum chamber and allowing the pressure in the cell to reach a new equilibrium, pumped only by non-evaporable getters. Under these conditions we evaporatively cooled an initial ensemble of $\sim 10^6$ $^6$Li atoms, prepared in the glass cell (see Supplemental Material), just beneath the $83.2$ mT Feshbach resonance to a final spin-balanced population of $N=3.5\times10^4$ total atoms. For this number of atoms, the Fermi energy  is $E_F=k_B\times 1.1(1)\ \mu K$, computed from the 3D density of states of our numerically-modeled trap. The ring and sheet powers were $P_r=1.3$ mW and $P_s=50$ mW, respectively. The axial and radial sheet trapping frequencies were measured using a parametric heating technique, and cross-checked using our sheet beam optical model, to be $f_z=1.6(1)$ kHz and $f_s=41(3)$ Hz, respectively. Furthermore, our trap model predicts $E_F/V_{evap}=0.95$.

Next, to ensure atom loss was predominantly due to one-body background collisions, and not due to rethermalizing collisions (discussed later) or parametric heating via trap vibrations, we ramped the sheet immediately after evaporation to $2.5$ times the minimum sheet power. This in turn approximately halved the ratio $E_F/V_{evap}$ and increased the sheet trap frequencies by a factor of $\sqrt{2.5}$. We subsequently ramped the magnetic field adiabatically to 100 mT where the interaction parameter $1/k_F|a|\approx 1.0$. At this stage, $T/T_F\approx 0.03$. Here, we held the atoms in the trap for varying amounts of time and extracted the reduced temperature $k_BT/\mu$ by fitting the in-situ density profile of the halo. To do this, we used a model that accounts for dimensional crossover in the outer regions of the halo where $\mu(r)\sim \hbar\omega_z$, and assumes that $k_BT\lesssim \hbar\omega_z$, which was true even for the highest temperatures measured (see Supplemental Material).

\begin{figure}[t]
	\centering
	\includegraphics[width=\columnwidth]{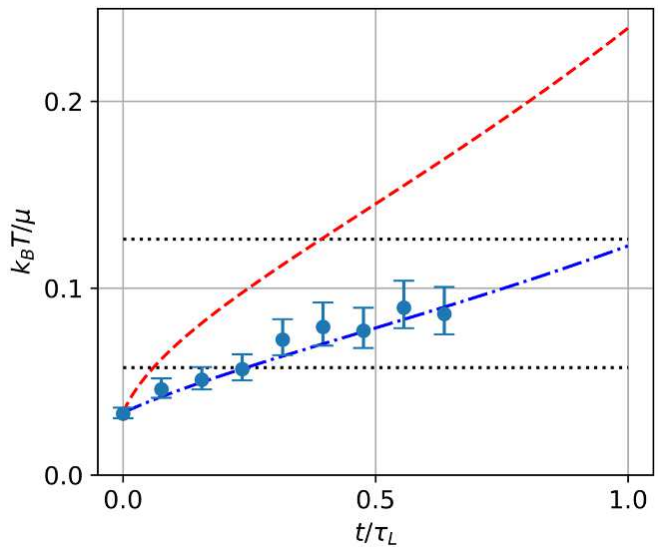}
	\caption{Reduced temperature versus time for an ideal Fermi gas in our trap potential. Blue circles are experimental data for an over-filled ring dimple with $N = 3.5\times 10^4$ atoms at $T/T_F=0.03$. The blue dash-dotted line is the temperature predicted by the model described in the text. The red dashed curve is the model's predicted temperature for $N=2.7\times 10^3$ atoms in the same potential, which just barely fills the ring. The black horizontal dotted lines show the threshold reduced temperature $(k_BT/\mu)_{max}$ required for pairing when $1/k_F|a|=1.0$ (lower) and $1/k_F|a|=0.5$ (upper).}
    \label{fig:heatingrateextrapolated}
\end{figure}

Fig.~\ref{fig:heatingrateextrapolated} shows the reduced temperatures measured in this configuration for different holding times. For comparison, we also plot the predicted temperature profile obtained by numerical integration of the heating rate equation \eqref{heatingraterevised}, for the initial conditions, trap parameters, and vacuum lifetime in the experiment. The theory and measurements agree to within the error shown in Fig.~\ref{fig:heatingrateextrapolated}, which was estimated from temperature fits using the upper and lower bounds of the sheet radial trap frequency, which is the dominant source of uncertainty. Fig.~\ref{fig:heatingrateextrapolated} also shows the predicted temperature increase for $2.7\times10^3$ atoms in this potential, which does not quite fill up the ring. The effects of hole-heating on the system temperature are significantly greater for a bare ring due to the reduced heat capacity per particle. In either case, hole heating also sets a practical limit on the lowest achievable $T/T_F$, due to both the finite state preparation time used in the experiment and the balance between the thermalization rate and hole-heating rate. This window of preparation time is narrower in the bare-ring configuration however, since the heating rate is roughly twice that of the ring-dimple for all hold times shown in Fig.~\ref{fig:heatingrateextrapolated}.

We now briefly draw connection to potential experiments performed in the BCS limit, which typically rely on maintaining temperatures below the pairing critical temperature. We start with Gor'kov's critical temperature prediction ~\cite{Gorkov1961}, $k_BT_c\approx0.277\mu\exp(-\pi\lambda/2)$, with $\lambda=1/k_F|a|$ the interaction parameter. Equating this expression to $k_BT$ gives the threshold reduced temperature below which BCS pairing (at the potential minimum) can occur for a given $\lambda$, i.e. $(k_BT/\mu)_{max}\equiv 0.277\exp(-\pi\lambda/2)$. We show these threshold reduced temperatures in Fig.~\ref{fig:heatingrateextrapolated} for $\lambda=1.0$ and $\lambda=0.5$. Clearly, a ring-dimple configuration can offer a substantially larger window of time to perform BCS-limit experiments compared to a bare-ring configuration. This could be especially important in an experimental apparatus with limited vacuum lifetime, or for experiments attempting to probe increasing $1/k_F|a|$ limits.

Because our temperature measurement method relies on fitting the density profile of the halo, some other method of thermometry would be required to measure the heating rate of a bare ring (see, for example, Refs. ~\cite{Chevy2022,Baillie2010}). These alternate techniques are likely to be more complex and involve more potential sources of error than extracting the temperature directly from a fit to the halo, however. The utility that a halo offers for temperature measurements in these kinds of fermionic systems should not be overlooked.

\textit{Thermalization and Loss.} In the idealized scenario described above, the equilibrium state after the ejection of an atom by a background collision is still a mostly-filled Fermi sea, and subsequent elastic collisions within the system will tend to repopulate the empty state. This typically occurs when two atoms at the Fermi level scatter (Pauli blocking suppresses scattering in the Fermi sea), one drops in energy to fill the empty state and the other is promoted to an energy $\epsilon\geq E_F$, in something like an Auger process. If $\epsilon\geq V_{evap}$, the excited atom can escape from the trap, and the new equilibrium is a filled Fermi sea with $N-2$ atoms. This loss of an additional atom always occurs (at $T=0$) for $V_{evap}=E_F$, and the probability decreases to zero when $V_{evap}=2E_F$, since the maximum scattering energy is $2E_F$.

More generally, the additional loss above the background rate will depend on other quantities that may include the ratio $E_F/V_{evap}$, the elastic collision rate, temperature, and conserved quantities pertaining to the trap potential. Experimentally we observed that the initial loss rate was three times the background rate when we did not increase the sheet depth after evaporation. This can occur if fermions scattered via the Auger process (with energy up to $2E_F$) scatter off another fermion and in turn excite another fermion to an energy above the Fermi level  (up to $1.5E_F$), which can also escape if its energy is above $V_{evap}$. This process can repeat if one or both of these atoms remain in the trap long enough. Thus, a single background collision in our ring-dimple trap may seed a cascade of energy from a single highly excited ``Auger" fermion to a state of many weakly excited fermions above the Fermi level, some of which may escape the trap. A non-trivial trap geometry can make the re-equilibration dynamics quite complicated, but qualitatively we would expect modifications to the loss and heating rates especially for $E_F\approx V_{evap}$. In this case the system would typically experience increased initial loss, with high energy atoms being lost from the trap, keeping the temperature low but causing the Fermi energy to drop rapidly. The loss rate would also become time dependent, asymptotically approaching the vacuum-limited loss rate as the Fermi energy drops well below the evaporation depth. These re-equilibration dynamics in ring-shaped systems are interesting in their own right, and further experimental and theoretical investigation is warranted.

\textit{Conclusion} We have demonstrated that one-body loss in a ring-shaped ensemble of ultracold fermions causes heating. We predicted the rate of temperature rise using a model that accounted for hole-induced heating, and have argued that this heating can be reduced by a particular choice of trap configuration. In particular, maintaining a large, dilute atomic background in contact with the ring helps to dissipate energy imparted into the ensemble via background collisions, which in turn keeps the temperature low for longer periods of time. A high quality vacuum is still essential to ensure that timescales for heating are long enough to permit low-energy long-wavelength experiments on superfluids with low critical temperatures, but there are clear advantages to consider forgoing the simplicity of a bare-ring configuration in favor of the more complex but useful ring-dimple configuration.

This paper has focused on heating rates in the weakly attractive BCS limit, but it is important to emphasize that coherent detection of supercurrents in that limit by matter-wave interference has so far only been possible after ramping interactions to the BEC regime before ballistic expansion. Adiabatic compression occurs during such a ramp, however~\cite{Carr2004}, and we have seen that if the initial BCS temperature is too high, the heating incurred during the ramp can drive the system above the critical temperature for molecular condensation. Finding ways of mitigating or circumventing this problem will likely be an important part of future experimental efforts.

\textit{Acknowledgements}
We thank Roberto Onofrio for insightful discussions and careful reading of the manuscript. This work was supported by the National Science Foundation (Grant No. 2046097).

\end{document}


\title{Supplemental Material: Mitigating Heating of Degenerate Fermions in a Ring-Dimple Atomic Trap }

\author{Daniel G. Allman}
\author{Parth Sabharwal}
\author{Kevin C. Wright}

\email{Daniel.G.Allman.GR@dartmouth.edu}

\affiliation{Department of Physics and Astronomy, Dartmouth College, 6127 Wilder Laboratory, Hanover NH 03766, USA}

\maketitle

\subsection{Experimental Apparatus and State Preparation}
Our experimental apparatus is designed to produce ultracold gases of $^6$Li atoms in highly configurable optical dipole traps. After initial cooling and capture using a 2DMOT, 3DMOT, and crossed-beam optical dipole trap, we use a movable optical trap to transport $10^6$ atoms to the center of a glass cell located between vertically oriented confocal objective lenses (NA$=0.3$). The lower objective projects the ring-pattern potential generated from a pair of axicons onto the atoms in the cell while the upper objective is used for imaging. Magnet coils surrounding the cell can generate a nearly uniform magnetic field of up to 108 mT. We prepare the $^6$Li atoms in an equal spin mixture of the two lowest energy spin states, for which there is a broad Feshbach resonance at 83.2 mT, using a RF field sweep. For more detailed information see Supplemental Material of ~\cite{CaiPersistentPRL2022}

\subsection{Temperature Measurement} We obtained an estimate of the system temperature by fitting an appropriate theoretical model to the column density of the halo. In this region, we can approximate the potential $V(r,z)\approx V_{0,s} + \frac{m\omega_z^2z^2}{2} + V_{sheet}(r)$ where $V_{0,s}$ is the potential energy offset of the full trap at the origin and $V_{sheet}(r)$ is the azimuthally-symmetric sheet potential, with $V_{sheet}(r=0)\equiv 0$. To allow for the possibility of mixed dimensionality in our description of the density profile, we quantize the vertical motion to harmonic oscillator levels, while treating the radial motion semi-classically. This procedure is similar in spirit to the theoretical treatment of a quantum well in solid state systems. In this way, we may write a hybrid description of the density of states
\begin{equation}\label{hybridDoS}
    g_j(E)=\frac{s}{(2\pi\hbar)^2}\int d^2rd^2p\ \delta\left(E-\frac{p^2}{2m}-\hbar\omega_zj-V_r(r)\right)
\end{equation}
which represents the density of available states in the $j^{th}$ axial harmonic oscillator level ($j=0,1,...)$, for a system with $s$ spin degrees of freedom. We have defined $V_r(r)=V_{0,s}+V_{sheet}(r)$, where the zero point energy of the axial motion has been set to zero. This is consistent with the convention that the Fermi energy of a zero-density ensemble is taken to be zero. Again, we assume states with $E>V_{evap}$ do not contribute to the density of states. Integrating over momenta and summing over $j$, we identify the local density of states 
\begin{equation}\label{localHybridDoS}
    g(r;E)=s\frac{m}{2\pi\hbar^2}\left\lceil\frac{E-V_r(r)}{\hbar\omega_z}\right\rceil\Theta(E-V_r(r))\Theta(V_{evap}-E)
\end{equation}
where $\lceil x\rceil$ is the ceiling function. The column density $n_2(r)$ is found by integrating the Fermi-Dirac-weighted local density of states over $E$, giving
\begin{equation}\label{ncol}
\begin{split}
    n_2(r)&=\frac{s}{\lambda_T^2}\int_0^{\eta_c-\frac{q-q(r)}{\gamma}}dx\frac{\lceil x/\gamma\rceil}{e^{x-q(r)}+1} \\ 
    &=\frac{s}{\lambda_T^2}\sum_{j=0}^{\infty}\Biggl\{F_0[q(r)-\gamma j]\\
    &-F_0\left[q(r)-\gamma j + \frac{q-q(r)}{\gamma}-\eta_c\right]\Biggr\}
\end{split}
\end{equation}
where we have further defined $\lambda_T^2\equiv\frac{2\pi\hbar^2}{mk_BT}$,  $\gamma\equiv\frac{\hbar\omega_z}{k_BT}$, $q(r)\equiv\frac{\mu-V_r(r)}{k_BT}$, $\eta_c=\frac{V_{evap}}{\hbar\omega_z}$ and $F_0(x)=\log(1+e^x)$. We note, however, that typically $\eta_c\gtrsim 20$, and so the second term in the summation form of \eqref{ncol} may be neglected, and we will assume this approximation in the subsequent analysis. The integral form of \eqref{ncol} looks remarkably similar to the order $1$ Fermi-Dirac integral used to describe the 3D (column) density, except for the presence of the ceiling function in the integrand, which accounts for the discrete axial energy levels. This discreteness is blurred out if either $\gamma$ or $\gamma/q(r)$ is small compared to unity, which corresponds to the $3D$ limit. In this case, we can replace $\lceil x/\gamma\rceil$ with $x/\gamma$, and the resulting expression gives the proper integrated $3D$ column density
\begin{equation}\label{3Dlimit}
    n_2(r)\approx \frac{s}{\gamma\lambda_T^2}F_1(q(r))\ ; \ \ \ \ \ \gamma\ll 1\ \text{or}\ \gamma/q(r)\ll 1
\end{equation}
where $F_{\nu}(x)$ is the usual Fermi-Dirac integral of order $\nu$. Conversely, if $\gamma\gg 1$ and $\gamma/q(r)\gg 1$, we approach the $2D$ limit, and we may replace $\lceil x/\gamma\rceil$ with $1$, and the resulting column density gives the proper $2D$ density
\begin{equation}\label{2Dlimit}
    n_2(r)\approx\frac{s}{\lambda_T^2}F_0(q(r))\ ; \ \ \ \ \ \gamma\gg 1\ \text{and}\ \gamma/q(r)\gg 1
\end{equation}
At this point, we have not assumed a particular form for the radially symmetric sheet potential. If we do have knowledge of the sheet trap parameters, however, we may use them to eliminate a fit parameter from the fitting function. In our case, the sheet potential may be described by $V_{sheet}(r)=V_0(1-e^{-2r^2/w_s^2})$ where $V_0=\frac{m\omega_s^2w_s^2}{4}$ and $\omega_s$ and $w_s$ are the sheet radial angular trapping frequency and $1/e^2$ radius, respectively. We may therefore introduce $\eta\equiv V_0/(k_BT)$ to write $q(r)=q-\eta(1-e^{-2r^2/w_s^2})$ and subsequently eliminate $\gamma=\frac{\hbar\omega_z}{k_BT}=\frac{\hbar\omega_z}{V_0}\frac{V_0}{k_BT}\equiv\frac{\eta}{\mathcal{N}}$ as a free fit parameter, assuming $\mathcal{N}=V_0/(\hbar\omega_z)$ is a known, albeit potentially uncertain input. We rewrite the column density \eqref{ncol} as 
\begin{equation}\label{fullfitfunc}n_2(r)=n_{\infty}+n_0\frac{\sum_{j=0}^{j_{max}}F_0\left[q-\eta(1-e^{-2(\frac{r-r_0}{w_s})^2}+\frac{j}{\mathcal{N}})\right] }{\sum_{j=0}^{j_{max}}F_0\left(q-\frac{j\eta}{\mathcal{N}}\right)}
\end{equation}
where we have introduced $n_0$ as the column density at $r=r_0$ and allowed for a non-zero density offset $n_{\infty}$ and center shift $r_0$ in the density profile. $j_{max}$ is the number of terms to include in the sum, and typically doesn't need to be very large. In total, there are five free parameters $(q,\eta,n_0,n_{\infty},r_0)$ that can be extracted via a least square fitting routine. However, one should ensure that $n_{\infty}$, $n_0$ and $r_0$ are as tightly bound and accurate as possible. This is achieved via careful image processing and pre-fitting analysis of the density profiles. Furthermore, the guesses for the remaining $q$ and $\eta$ should be physically reasonable. Namely, $\eta\geq0$ and $q$ is not too negative if dealing with a presumed near-degenerate ensemble. 

It is also important to note that, for deeply degenerate Fermi gases, absolute temperature enters into the fit of the data only in the far dilute wings of the density distribution. The use of a hybrid fitting function \eqref{fullfitfunc} was motivated by this fact, and we found that a simple 3D Fermi-Dirac function consistently overestimated the density at large radii. We found empirically that the largest source of uncertainty in the temperature estimate is from our measurement of the weak radial trap frequency of the sheet beam, $f_s$.  Measurement noise introduces a much smaller uncertainty, and uncertainty in the axial trap frequency introduces a similarly small amount. Uncertainty in imaging beam parameters such as saturation intensity and polarization impurity will introduce systematic errors onto the temperature estimate. For the dilute halo atoms, error due to saturation effects may be neglected, and polarization purity is almost unity, so uncertainty due to it may be neglected. Finally, at large radii, the sheet becomes slightly elliptical, and this in turn causes a small systematic shift in measured temperatures to smaller values. This shift becomes more apparent in thermal ensembles where the extent of the atomic distribution into these elliptical regions is larger.

\subsection{Local Fermi Energy}
\begin{figure}[t!]
\centering
\includegraphics[width=20cm,height=10cm,keepaspectratio]{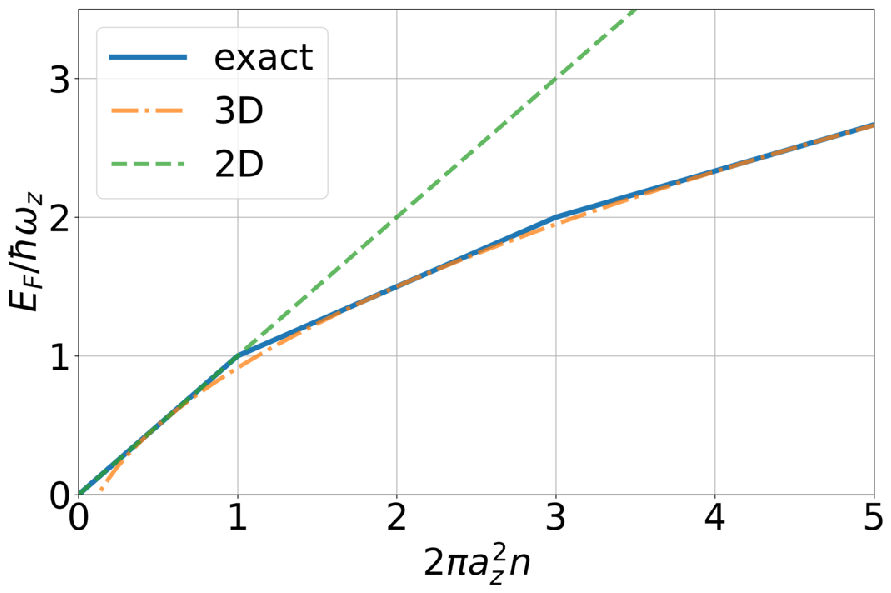}
\caption{Reduced Fermi energy as a function of reduced column density, with the $2D$ ($E_F/\hbar\omega_z\sim\tilde{n}$) and $3D$ ($E_F/\hbar\omega_z\sim\sqrt{2\tilde{n}}-1/2$) limiting behavior shown for comparison. The $2D$ prediction matches the exact prediction for $\tilde{n}\leq\tilde{n}_c\equiv 1$.}
\label{fig:FermiEnergyPlot}
\end{figure}
It is often insightful to study thermodynamic quantities locally within a local density approximation (LDA) framework. The local Fermi energy in particular is a useful energy scale to normalize i.e. the chemical potential and thermal energy with. Again, we would like to account for the possibility of a dimensional crossover in the halo, and expect deviations of the true local Fermi energy from its 3D equivalent at low densities. We define the local Fermi energy $E_F(r)$ by integrating \eqref{localHybridDoS} as follows:
\begin{equation}\label{EFlocal}
\begin{split}
    n(r)&\equiv\int_{V_r(r)}^{E_F(r)+V_r(r)}g(r;E)dE \\ 
    &=\frac{1}{2\pi a_z^2}\int_0^{x}dz\lceil z\rceil \\
    &=\frac{1}{2\pi a_z^2}(1+2+...+(j_c-1)+j_c\{x\})
\end{split}
\end{equation}
where $n(r)$ is the column density, $x\equiv E_F(r)/\hbar\omega_z$, $\{x\}$ denoting the fractional part of $x$, and $a_z^2\equiv\hbar/(m\omega_z)$. $j_c\equiv\lceil  x\rceil$ represents the number of populated axial levels. Note that this definition of the local Fermi energy gives $E_F(r)=0$ for $n(r)=0$, which is conventional. The sum forms a triangle number, and by defining $\tilde{n}=2\pi a_z^2n$, we may write (dropping the position label for now)
\begin{equation}\label{EFlocal2}
    \tilde{n}=j_c\left(j_c/2+\{x\}-1/2\right)
\end{equation}
This equation may be analytically inverted to find the reduced Fermi energy $x$ as a function of the reduced density $\tilde{n}$ as follows. First, we rearrange \eqref{EFlocal2}
\begin{equation}\label{EFlocalre}
    \{x\}-1/2=\frac{\tilde{n}}{j_c}-\frac{j_c}{2}
\end{equation}
Since the fractional part obeys $0\leq\{x\}\leq 1$, the L.H.S. is bound between $-1/2$ and $1/2$. Therefore, we seek solutions $j_c$ that obey $-1/2\leq \frac{\tilde{n}}{j_c}-\frac{j_c}{2}\leq 1/2$. Solving for $j_c$ at the endpoints gives rise to the constraint that 
\begin{equation}\label{constraint}
    \frac{-1+\sqrt{1+8\tilde{n}}}{2}\leq j_c \leq\frac{1+\sqrt{1+8\tilde{n}}}{2}
\end{equation}
for valid solutions to \eqref{EFlocalre}. As $j_c$ is an integer, and the region in \eqref{constraint} has a size of $1$ for all $\tilde{n}$, there must be exactly one $j_c$ in that region, whose value is simply given by $j_c=\lfloor \frac{1}{2}(1+\sqrt{1+8\tilde{n}})\rfloor$. Plugging $j_c$ into \eqref{EFlocalre}, and using $x=\lceil x\rceil +\{x\}-1=j_c+\{x\}-1$, the local Fermi energy in this hybrid picture may be written in closed, universal form as
\begin{equation}\label{EFexact}
    \frac{E_F(r)}{\hbar\omega_z}=\frac{\tilde{n}(r)}{\lfloor \frac{1}{2}(1+\sqrt{1+8\tilde{n}(r)})\rfloor}+\frac{\lfloor \frac{1}{2}(1+\sqrt{1+8\tilde{n}(r)})\rfloor}{2}-\frac{1}{2}
\end{equation}

 Fig.~\ref{fig:FermiEnergyPlot} shows the exact Fermi energy as a function of the $2D$ density. The Fermi energy reduces to the known results in the $2D$ $(0\leq\tilde{n}\leq 1$) and axially-integrated $3D$ $(\tilde{n}\gg 1$) limits, and correctly accounts for the discrete nature of the axial harmonic oscillator levels. It is important to remember that this description only holds \textit{locally} in the halo, and does not represent the global Fermi energy of our ring dimple trap.